\documentclass[referee]{raa}            
\usepackage{graphicx,times}             
\usepackage{natbib}
\usepackage{amssymb,amsmath}
\bibpunct{(}{)}{;}{a}{}{,}
\usepackage{threeparttable}
\usepackage{multirow}
\usepackage[figuresright]{rotating}
\usepackage{threeparttable}
\usepackage{enumerate}
\usepackage[pagebackref=true]{hyperref}
\usepackage{multirow}

\begin{document}

   \title{A study of 10 Rotating Radio Transients using Parkes radio telescope}

 \volnopage{ {\bf 20XX} Vol.\ {\bf X} No. {\bf XX}, 000--000}
   \setcounter{page}{1}

   \author{Xinhui Ren
   \inst{1,2,3}, Jingbo Wang\inst{3}, Wenming Yan\inst{1,4}, Jintao Xie
      \inst{5},  Shuangqiang Wang\inst{1}, Yirong Wen\inst{1,2,3}, Yong Xia\inst{1,2,3}
   }

   \institute{ Xinjiang Astronomical Observatory, Chinese Academy of Sciences, Urumqi, Xinjiang 830011, People's Republic of China\\
        \and
             University of Chinese Academy of Sciences, Beijing 100049, People's Republic of China\\
	\and
            Institute of Optoelectronic Technology, Lishui University, Lishui, Zhejiang, 323000, People's Republic of China; {\it 1983wangjingbo@163.com}\\
        \and
            Xinjiang Key Laboratory of Radio Astrophysics, Urumqi, Xinjiang, 830011, People's Republic of China\\
        \and
            Research Center for Intelligent Computing Platforms, Zhejiang Laboratory, Hangzhou, Zhejiang 311100, China\\
\vs \no
   {\small Received 20XX Month Day; accepted 20XX Month Day}
}

\abstract{Rotating Radio Transients (RRATs) are a relatively new subclass of pulsars that emit detectable radio bursts sporadically. We conducted an analysis of 10 RRATs observed using the Parkes telescope, with 8 of these observed via the Ultra-Wideband Receiver. 
We measured the burst rate and produced integrated profiles spanning multiple frequency bands for 3 RRATs. We also conducted a spectral analysis on both integrated pulses and individual pulses of 3 RRATs.  
All of their integrated pulses follow a simple power law, consistent with the known range of pulsar spectral indices. 
Their average spectral indices of single pulses are -0.9, -1.2, and -1.0 respectively, which are within the known range of pulsar spectral indices.   
Additionally, we find that the spreads of single-pulse spectral indices for these RRATs (ranging from -3.5 to +0.5) are narrower compared to what has been observed in other RRATs \citep[][]{2018ApJ...866..152S,2022ApJ...940L..21X}. It is notable that the average spectral index and scatter of single pulses are both relatively small. For the remaining 5 RRATs observed at the UWL receiver, we also provided the upper limits on fluence and flux density. In addition, we obtained the timing solution of PSR J1709-43. 
Our analysis shows that PSRs J1919+1745, J1709-43 and J1649-4653 are potentially nulling pulsars or weak pulsars with sparse strong pulses. 
\keywords{pulsars: general --- stars: neutron
}
}

   \authorrunning{Xinhui Ren et al.: A study of 10 RRATs}            
   \titlerunning{Xinhui Ren et al.: A study of 10 RRATs}  

   \maketitle
%
%
\section{Introduction}           
\label{sect:intro}

Pulsars serve as the cosmic lighthouses and can be a powerful tool for numerous astronomical studies. They are commonly utilized in testing General Relativity \citep[GR; e.g.][]{2006Sci...314...97K,2013Sci...340..448A}. Pulsars are frequently timed for use in Pulsar Timing Arrays (PTAs) due to their stable spin periods \citep[e.g.][]{2010CQGra..27h4013H,2013PASA...30...17M} and for probing the nature of the interstellar medium \citep[ISM;][]{2018ApJS..234...11H}. While the majority of known pulsars are discovered through periodicity-based searches \citep{2012hpa..book.....L}, Rotating Radio Transients (RRATs) , a subclass of canonical pulsars, were detected through single-pulse searches instead of the standard Fourier domain searches or conventional folding techniques \citep{2006Natur.439..817M,2011BASI...39..333K}.

RRATs are notable for their sporadic emissions, with only a few pulses being detected each hour\footnote{http://astro.phys.wvu.edu/rratalog/} \citep[e.g.][]{2011MNRAS.415.3065K}. Another intriguing characteristic of RRATs is their similarity in radio emissions to Fast Radio Bursts (FRBs), which are extremely bright and manifest as millisecond-duration burst events. Single-instance RRAT pulses appear indistinguishable from FRB pulses. Dispersion Measure (DM) is used to quantify the time delay caused by the interstellar and intergalactic medium. Generally, FRBs tend to have larger DM values due to their extragalactic origin \citep{2018MNRAS.475.1427B}, while RRATs tend to exhibit relatively lower DM values \citep{2023MNRAS.524.5132D}. Previous studies have attempted to identify potentially misclassified FRBs as RRATs, especially when they have only been detected through a single-pulse approach \citep{2016MNRAS.459.1360K,2016arXiv160806952R}. It has been suggested that magnetars, another subsection of neutron stars with high magnetic fields \citep{2017ARA&A..55..261K}, may be responsible for high-energy phenomena such as FRBs \citep{2020Natur.587...54C,2020Natur.587...59B}. Consequently, discovering more RRATs and magnetars can provide insights into the distribution of Galactic pulsars and potentially help characterize the populations of RRATs, FRBs, and magnetars.

RRATs are Galactic pulsars characterized by extremely variable emission of single pulses. They emit individual pulses but then exhibit long periods with no detectable emission. However, only several telescopes with high sensitivity are thought to be suitable to detect such intermittent emissions and obtain sufficient single pulses and periodic emission statistics.  This inherent challenge in RRAT detection has resulted in incomplete knowledge regarding the fraction of RRATs that exhibit nulling behavior. Currently, a large number of RRATs lack a precisely measured rotation period, approximately two-thirds have not had their burst rates determined due to a lack of follow-up observations. This lack of data significantly hinders our ability to properly characterize these sources, as highlighted in the study by \cite{2023arXiv230212661M}. Consequently, in terms of RRATs, it is necessary to conduct long-term monitoring and timing observations.

The sporadic emission mechanism of RRATs remains unknown, primarily due to the inherent difficulties associated with their detection. Despite the discovery of nearly 170 RRATs as of 2023 \citep[e.g.][]{2018A&A...618A..70T,2021ApJ...922...43G}, the elusive nature of RRATs has posed significant challenges in unraveling their emission behavior. Since their initial discovery, several models have been put forward to explain these sporadic phenomena. Some examples include: a normal pulsar exhibiting extreme nulling \citep[e.g.][]{2007MNRAS.377.1383W,2010MNRAS.402..855B}, fallback of supernova material \citep{2006ApJ...646L.139L}, trapped plasma being released from radiation belts \citep{2007MNRAS.378.1481L}, and interference of asteroidal or circumpulsar debris \citep{2008ApJ...682.1152C}. Alternatively, it could be related to mechanisms within the pulsar magnetosphere \citep[e.g.][]{2010MNRAS.408L..41T,2012ApJ...746L..24L,2014MNRAS.437..262M}. Efforts have been devoted to understanding this intriguing phenomenon. To establish potential connections between RRATs and normal pulsars, \cite{2023arXiv230317279Z} conducted an in-depth analysis of 76 Galactic rotating radio transients and their enigmatic emission patterns. The study concluded that RRATs predominantly are weaker pulsars with a few strong pulses or extreme nulling pulsars.

In this paper, we present observations of 10 RRATs conducted at the Parkes 64m radio telescope, utilizing the ultra-wide-bandwidth low-frequency (UWL) receiver, which covers the continuous frequency range of 704-4032 MHz \citep{2020PASA...37...12H}. This receiver system provides unprecedented broadband information, including valuable polarization data, that plays a pivotal role in advancing our understanding of radiation and various phenomena within the intervening interstellar medium (ISM), such as scintillation and dispersion measure (DM) variability \citep[e.g.][]{2015ApJ...809...67K,2020Galax...8...53H}. Apart from its large fractional bandwidth, it also has a low system temperature, allowing it it to operate effectively even in the presence of strong mobile phone transmissions \citep{2020PASA...37...12H}. In recent years, the UWL receiver has played a crucial role in numerous scientific projects related to high-precision pulsar timing, the broadband analysis of pulsar profiles, and the detection of new pulsars and transient sources. Additionally, we analyzed the radiation characteristics of two RRATs observed using other receivers at the Parkes telescope.


\section{Observations}
\label{sect:Obs}

A majority of pulsar observations on the Parkes Radio Telescope archive can be obtained at the CSIRO pulsar data archive since 1991. Data sets for a large number of observations will be publicly available after an 18-month embargo\footnote{https://data.csiro.au/domain/atnf} \citep{2011PASA...28..202H}. The observations from the Parkes telescope have been obtained under project codes, mainly P1016 - Instant GRRATificiation. The RRATs in our sample were observed using the Multibeam, H-OH, 10/50 and UWL receivers \citep[e.g.][]{2013PASA...30...17M,2020PASA...37...12H}. Most of these observations were conducted using the UWL receiver system, with 3328 MHz of bandwidth centered on 2368 MHz and folded in real time into 1024 phase bins. The receiver system provides continuous frequency coverage from 704 to 4032 MHz and has a low temperature \citep{2020PASA...37...12H}. PSRs J1649-4653 and J1709-43 were observed by the Multibeam receiver with 256 MHz of bandwidth divided into 1024 frequency channels and folded in real time into 1024 phase bins. Additionally, the observed bandwidth is normally 512 MHz for the H-OH receiver and 1024 MHz for the 10/50 receiver. All data were recorded with the PSRFITS data format \citep{2004PASA...21..302H}. In the observing course, we used 5 backend systems, which included Medusa and the Parkes Digital FilterBank systems \citep[PDFB1,PDFB2,PDFB3, PDFB4;][]{2006MNRAS.369..655H}.

\begin{table}
\bc
\begin{minipage}[]{100mm}
\caption[]{Observing Parameters for 10 RRATs.\label{tab1}}\end{minipage}
\setlength{\tabcolsep}{1pt}
\small
 \begin{tabular}{lccccccccc}
  \hline\noalign{\smallskip}

 PSR      &  Duration     & $N_p$& $N_b$    & Burst Rate&Backend&Frequency&Bandwidth &Receiver&Mode\\
       & (h)      & (pulses)       & (pulses)             & ($h^{-1}$)& &(MHz)&(MHz)&\\
\hline

J0627+16      &   0.7  &  1140  &  -  &   - & MEDUSA&2368&3328&UWL &Search\\
J0628+0909      &   1.0  &  3001  &  61  &   60& MEDUSA&2368&3328& UWL&Search \\
J1850+15      &   0.7 &  1867  &  -  &   - & MEDUSA&2368&3328& UWL &Search\\
J1909+0641      &   2.0  &  9730  &  116  &   58 & MEDUSA&2368&3328& UWL&Search \\
J1913+1330      &   0.9  &  3459  &  -  &   -  & MEDUSA&2368&3328& UWL&Search\\
J1919+1745      &   1.0  &  1737  &  835  &   847 & MEDUSA&2368&3328& UWL&Search \\
J1928+15      &   0.3  &  3012  &  -  &   -  & MEDUSA&2368&3328& UWL&Search\\
J1946+24      &   0.3  &  257  &  -  &   -  & MEDUSA&2368&3328& UWL&Search\\

\noalign{\smallskip}\hline
~\\
\end{tabular}

\begin{tabular}{lccccccc}

  \hline\noalign{\smallskip}

PSR               &  Time Span          &Sub-int Length & Backend & Frequency& Bandwidth& Receiver&Mode\\
              &(Years)  & (s)   &       & (MHz)    &    (MHz)&\\
\hline

J1649-4653      &   10.8  & 10/20/30/60&  PDFB1/2/3/4    &   732/1369/3094  &    256/1024 &H-OH/MULTI/10/50CM &Fold\\
J1709-43      &   7.2 &10/20/30/50 &  PDFB3/4   &   845/1390/3380  &    
64/256/512&H-OH/MULTI/10/50CM &Fold\\

  \noalign{\smallskip}\hline
\end{tabular}
\ec

\tablecomments{\textwidth}{We have listed the observing parameters for 10 RRATs,but only PSRs J0628+0909, J1909+0641, J1919+1745 were measured the burst rate. $N_p$ represents the number of total pulses observed, while $N_b$ is the number of burst pulses. The sub-int length represents the sub-integration length.}
\end{table}

We investigated 10 RRATs in the CSIRO pulsar data archive and all observation data are publicly available as of 2021. The observing data for the UWL receiver are in search mode, while the data for the H-OH, Multibeam and 10/50 receivers are in fold mode. A summary of the observations is given in Table~\ref{tab1}, which lists the fundamental information of the telescope and the project. The search-mode data sets were subsequently folded by {\sc dspsr} \citep{2011PASA...28....1V}, at the rotation period and DM of the RRATs acquired using {\sc psrcat} \citep{2005AJ....129.1993M}. The {\sc paz} tool in {\sc psrchive} \citep{2004PASA...21..302H} was used to eliminate narrowband and
impulsive radio frequency interference (RFI). To further mitigate of potential RFI contamination, we utilized the {\sc pfits\_zapprofile} tool in  {\sc pfits} to process all the data. {\sc pfits} is a software package designed for reading, manipulating, and processing PSRFITS format pulsar astronomy data of both search mode and fold mode\footnote{https://bitbucket.csiro.au/projects/PSRSOFT/repos/pfits/browse}. Flux density calibration was made through the observation of the radio galaxy Hydra~A (3C 218), where the calibration method can be found in \citep{2020bda..book..165L}. The calibration data are then applied to the RRATs observations using the {\sc pac} tool to flatten the bandpass and transform the polarisation products to Stokes parameters. To transform the measured intensities to absolute flux densities, we used {\sc paas} to form noise-free standard templates from observations and then used {\sc psrflux} to obtain the flux density. In order to obtain the spectral information of RRATs, we divided the frequency bands into several parts and utilized the {\sc pulsar\_spectra} software to find the best-fitting model and produce publication-quality plots \citep{2022PASA...39...56S}.

\section{Analysis}
\label{sect:data}

\subsection{Identification of Single Pulse}

The first step in our analysis is to discern the pulses originating from the RRATs. The intermittent nature of RRAT emissions renders Fourier-based techniques or conventional folding search algorithms impractical. Following the method of \citet{2022ApJ...940L..21X}, we employ the single-pulse search method to specifically target individual pulses surpassing a predefined signal-to-noise ratio (S/N) threshold. 
Generally, a single-pulse phase can be segmented into two components: the on- pulse region and the off-pulse region. The off-pulse energy is commonly characterized by a Gaussian distribution.
We set a minimum detection threshold of 5$\sigma$ for each pulse, where $\sigma$ represents the standard deviation of the off-pulse region. To  filter out RFI, we exclusively consider pulses whose times of arrival (ToAs) derived from the brightest pulse observed during an observation fall within a 5\% deviation from the expected phase. For the purpose of minimizing potential contamination from RFI, each pulse was visually inspected, including an examination of its frequency-phase plots. The total number of detected pulses, the count of filtered pulses and the burst rates for all our target RRATs are listed in Table~\ref{tab1}. We only detected single-pulses in 3 RRATs (PSRs J1919+1745, J1909+0641 and J0628+0909).

\begin{figure}[h]
    \centering

    \begin{minipage}[t]{1.0\linewidth}
    \centering
        \begin{tabular}{@{\extracolsep{\fill}}c@{}c@{}c@{\extracolsep{\fill}}}
            \includegraphics[width=0.335\linewidth]{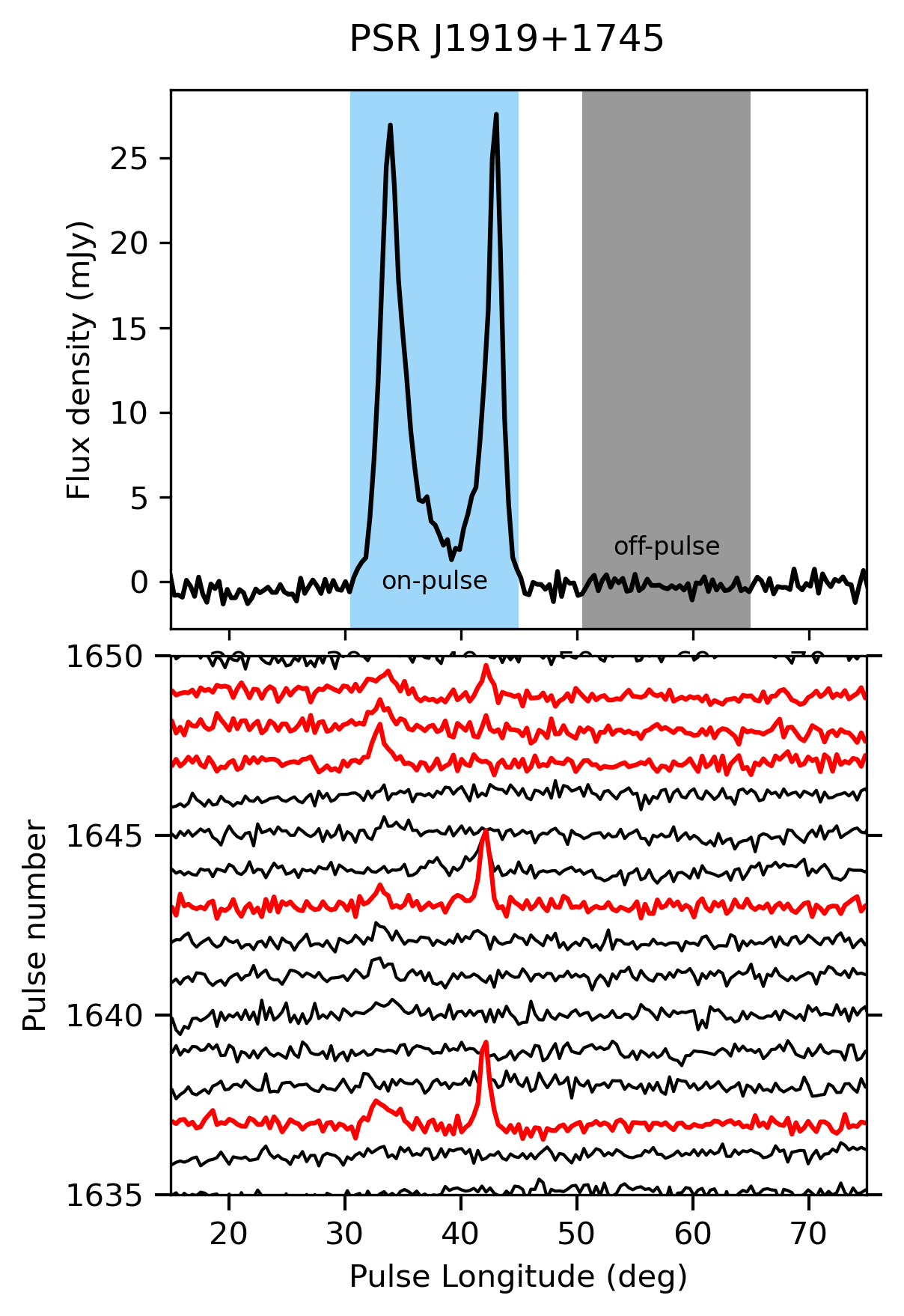}&
            \includegraphics[width=0.335\linewidth]{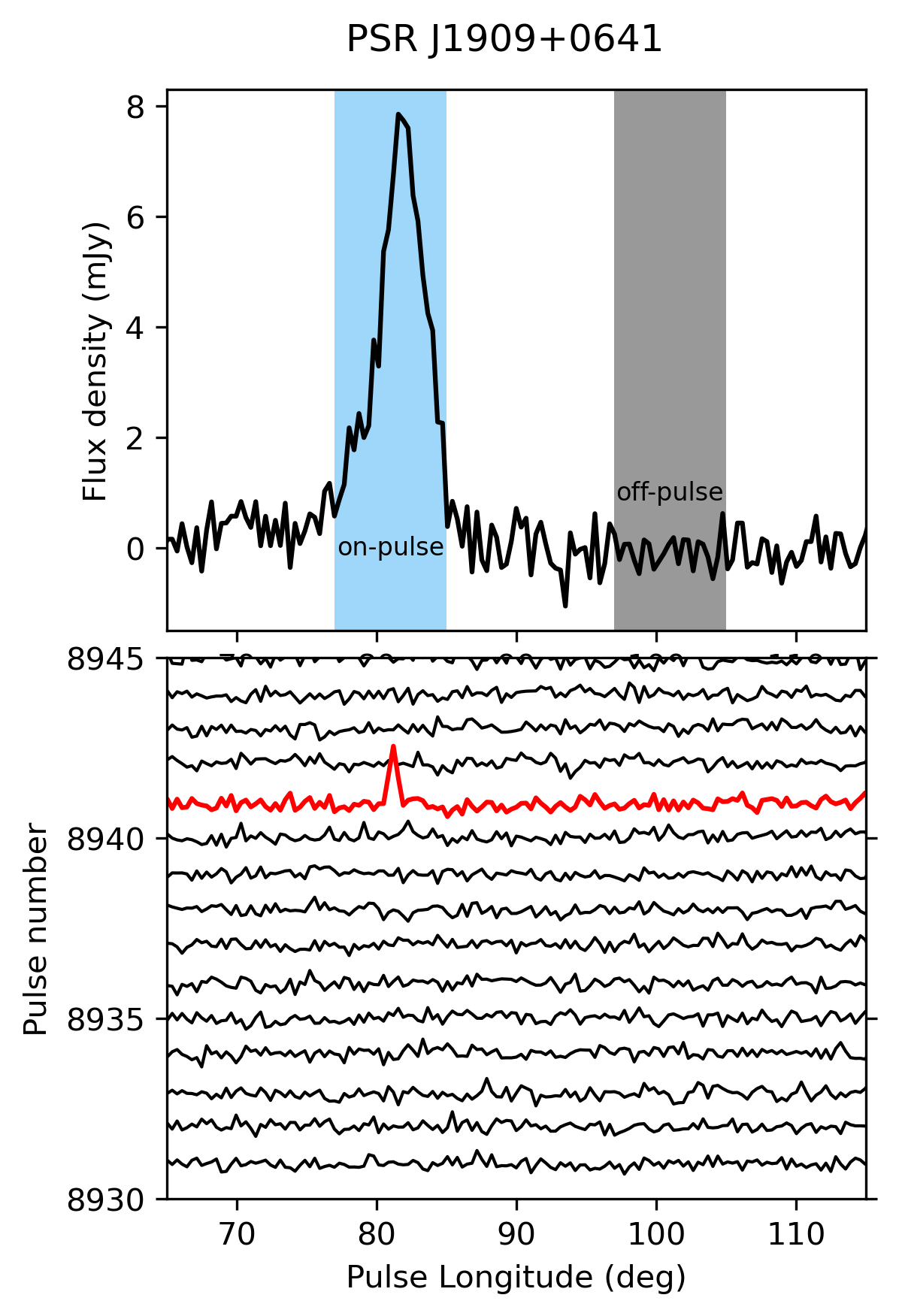}&
            \includegraphics[width=0.335\linewidth]{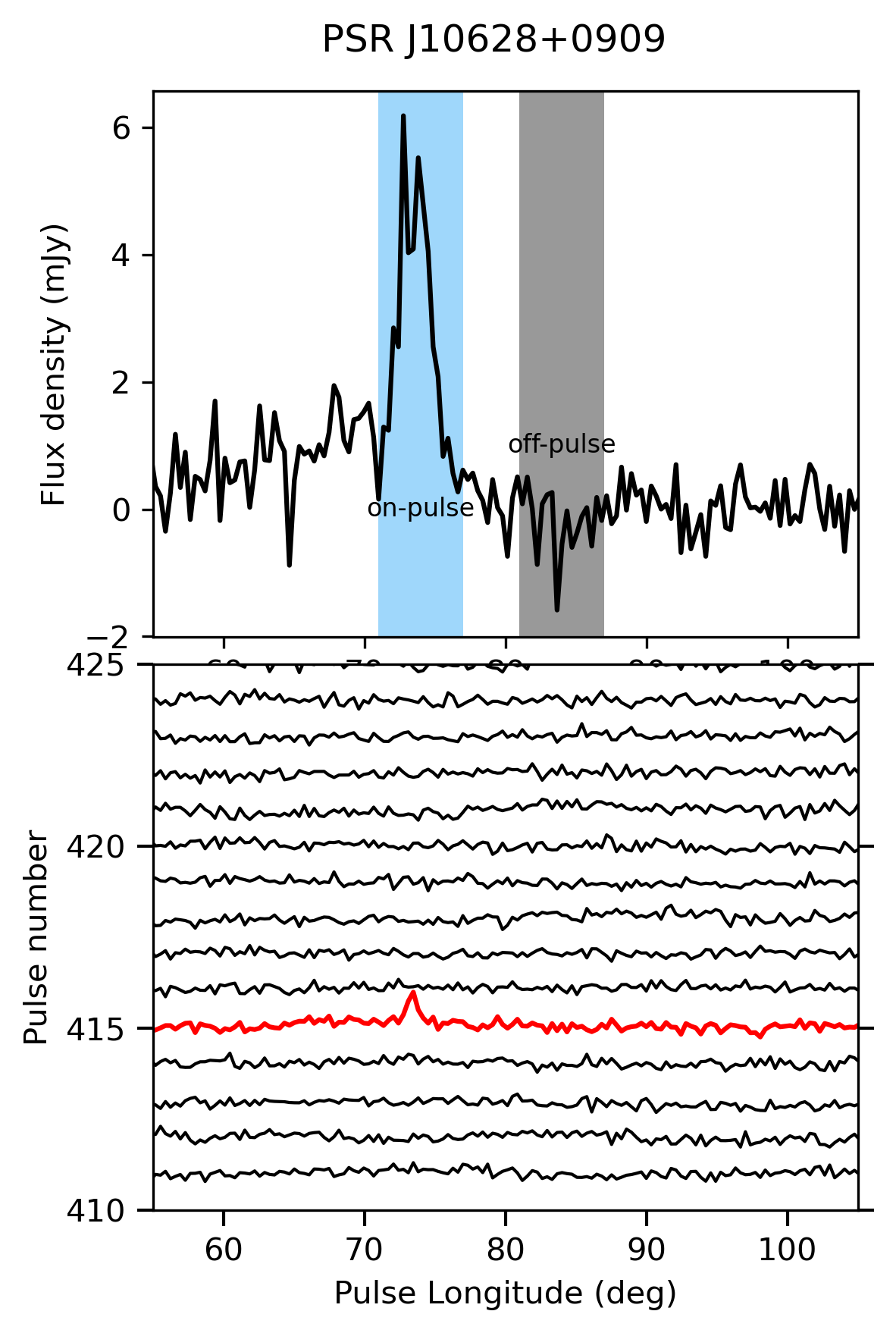}\\
           \\
        \end{tabular}
    \end{minipage}
    
    \caption{Integrated proflies and partial pulse sequences of 3 RRATs.}
    \label{sequence}
 \end{figure}

The integrated profiles of the entire observations for these 3 RRATs are shown in upper panel of Figure~\ref{sequence}, in which the on-pulse and off-pulse regions are shown in filled blue and gray areas, respectively.
In the lower panel of Figure~\ref{sequence}, we showed single-pulse stack of 15 single pulses for each RRAT, in which the burst pulses are labeled in red solid line. 
It is clear to see that all of these 3 RRATs exhibit sporadic emissions. The burst rate of PSR J1919+1745 is 847 $h^{-1}$, which is much larger than that of both PSR J0628+0909 with burst rate of 60 $h^{-1}$ and PSR J1909+0641 with burst rate of 58 $h^{-1}$. 

\subsection{Profile}

Flux calibration procedure was applied to all of the observational data. For the search-mode data, we conducted observations on each RRAT either once or twice, with each observation lasting approximately one hour. To manage individual file sizes, the data was automatically divided into multiple files. We utilized all observation files to generate a cumulative pulse profile, aiming to enhance the signal-to-noise ratio. For the fold-mode data, our analysis focused only on the integrated profile of PSR J1709-43. The sub-integration lengths across 103 observations varied between 10, 20, 30, or 50. To analyze the frequency-dependent characteristics of the integrated pulse profile, we utilized single observational data acquired from three frequencies: 732 MHz, 1369 MHz, and 3094 MHz. Each observation file contained 30 sub-integrations, pulse profiles were observed across all the sub-integrations. Therefore, the sub-integration length across observations did not exert an significant influence on the integrated profile.

\subsection{Flux Density Measurements and Spectral Properties}

We utilized the {\sc pulsar\_spectra} software package developed by \cite{2022PASA...39...56S}, which uses the robust statistical techniques to decide on the optimal fitting model and calculate the corresponding spectral parameters. The publicly available software package presents a comprehensive array of tools dedicated to the systematic cataloging of pulsar flux density and automated spectral fitting. The use of this software facilitates the identification of the most appropriate spectral model. Notably, the {\sc python}-based software implements the 5 spectral models, including the Simple power law, Broken power law, Log-parabolic spectrum, Power law with high-frequency cut-off, and Power law with low-frequency turnover.
They are adequate for describing the spectra of the vast majority of pulsars, with the simple and broken power-law models being the most widely employed. 
Although these models are morphological, the spectral index may potentially be associated with other parameters of pulsars. The spectral fitting routine implements the method illustrated in \cite{2018MNRAS.473.4436J}. To compare the 5 models, the Akaike information criterion (AIC) was used as a comparison metric, evaluating how much information the model maintains about the data without overfitting. The model leading to the lowest AIC is considered the most accurate in describing the pulsar's spectra. A more comprehensive description of how to use the {\sc pulsar\_spectra} software package can be found in \cite{2022PASA...39...56S}. We presented the results of spectral fitting from integrated pulses of several RRATs. Furthermore, we applied a similar spectral analysis procedure to individual pulses with high signal-to-noise ratio.

\section{Results}

\subsection{Search-mode data}

The radio emission from pulsars is affected by propagation through the ISM in both frequency and time domains, spanning durations from seconds to several hours. This phenomenon is recognized as diffractive scintillation. 
We estimate the scattering time, $\tau_s$ at a reference frequency of 1400 MHz using the empirical relationship to DM  acquired by\cite{2021MNRAS.501.4490K}: 
\begin{equation}
    \tau_s = 1.2\times10^{-5}DM^{2.2}(1.0+0.00194DM^2).
\end{equation}

For the wide frequency band of the UWL receiver, 
we applied the typical relation $\tau_\nu$ $\propto$ $\nu^{-4}$ \citep{1972AuJPh..25..759K} to estimate the scattering time at 950 MHz and 3500 MHz.

The value of scintillation bandwidth in MHz, $\Delta\nu_d$, is given by \cite{1998ApJ...507..846C} as:
\begin{equation}
    \Delta\nu_d = \frac{1.16}{2\pi\tau_s}.
\end{equation}

We estimated the scintillation bandwidth for 3 RRATs at frequencies of 950 MHz, 1400 MHz, and 3500 MHz, as shown in Table~\ref{tab-scin}. The entire observation bandwidth was divided into several sub-bands. For PSRs J1919+1745 and J0628+0909, the minimum bandwidth of each sub-band is 100 MHz, while for PSR J1909+0641, it is 200 MHz.The estimated scintillation bandwidths for PSRs J1919+1745, J1909+0641, and J0628+0909 are all significantly smaller than the minimum bandwidth of each corresponding sub-band. Therefore, the impact of scintillation effects on flux density is deemed negligible in this study.

\begin{table}[h]
\bc
\centering
\caption{Scintillation bandwidth and minimum sub-band bandwidth of 3 RRATs.}\label{tab-scin}
\begin{tabular}{lcccccc}
\hline

PSR &DM&$\Delta\nu_{min}$&$\Delta\nu_{950}$&$\Delta\nu_{1400}$&$\Delta\nu_{3500}$\\
&(pc $cm^{-3}$)&(MHz)&(MHz)&(MHz)&(MHz)\\
\hline
J1919+1745&142.3&100&0.001&0.007&0.3\\
J1909+0641&36.7&200&0.3&1.5&58.6\\
J0628+0909&88.3&100&0.01&0.05&2.0\\
\hline
\end{tabular}
\ec
\tablecomments{\textwidth}{We have listed the observing parameters for PSRs J1919+1745, J1909+0641, J0628+0909. 
$\Delta\nu_{min}$ is the minimum channel bandwidth. $\Delta\nu_{950}$, $\Delta\nu_{1400}$, and $\Delta\nu_{3500}$ are the scintillation bandwidth response to 950 MHz, 1400 MHz, 3500 MHz, respectively \citep{1972AuJPh..25..759K}.}
\end{table}

\subsubsection{PSR J1919+1745}

\begin{figure}[h]
\centering
\includegraphics[width=9cm]{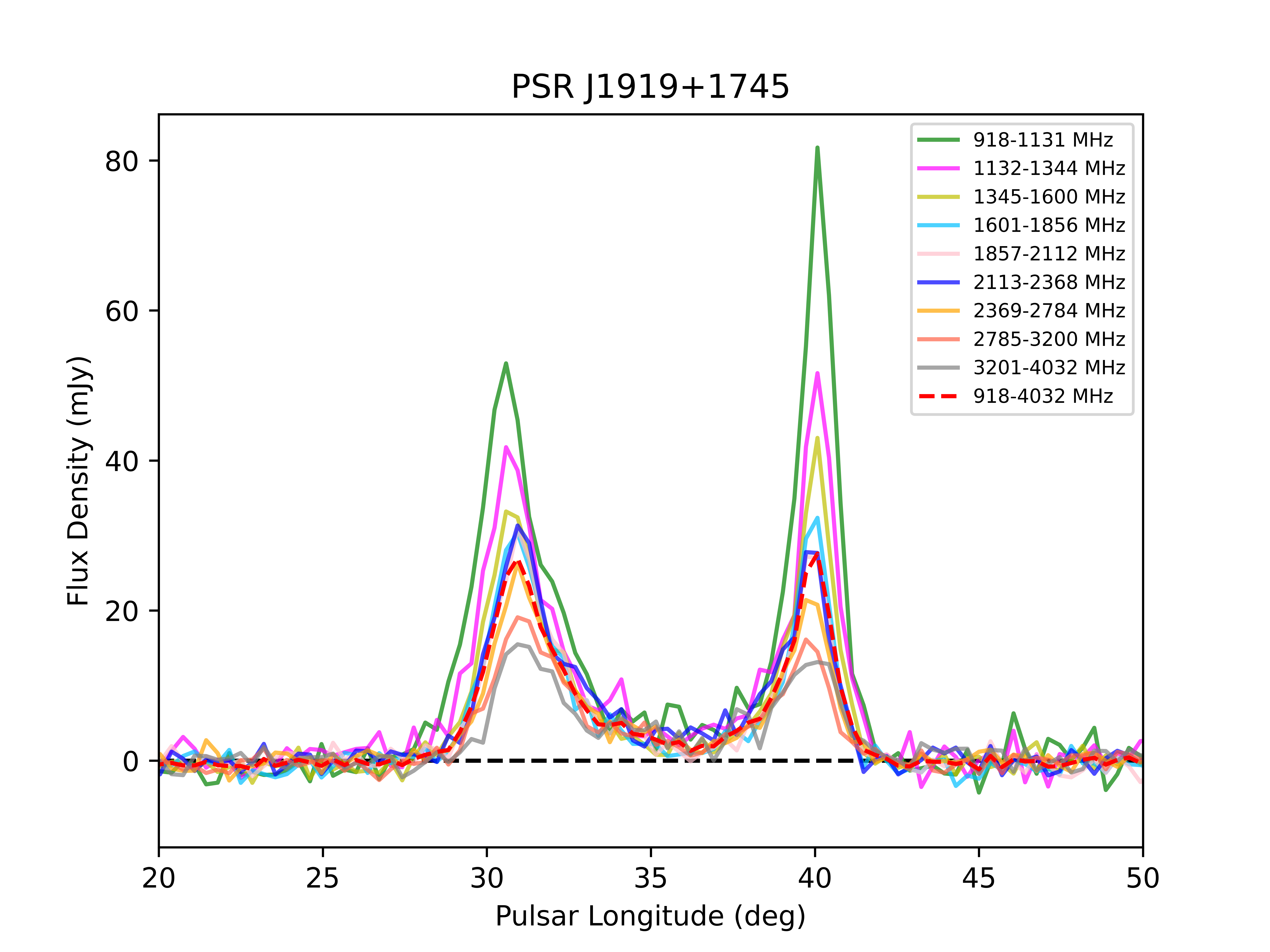}
\caption{Integrated profiles of PSR J1919+1745 across 9 frequency sub-bands.
\label{1919-9} }
\end{figure}

We have obtained 835 burst pulses, constituting approximately 48$\%$ of all detected single pulses. In Figure~\ref{1919-9}, we present the integrated profiles of the entire frequency band using 62 observations as well as those of nine sub-bands.
\begin{figure}[thbp!]
    \centering
    \begin{tabular}{@{\extracolsep{\fill}}c@{}c@{\extracolsep{\fill}}}
            \includegraphics[width=0.5\linewidth]{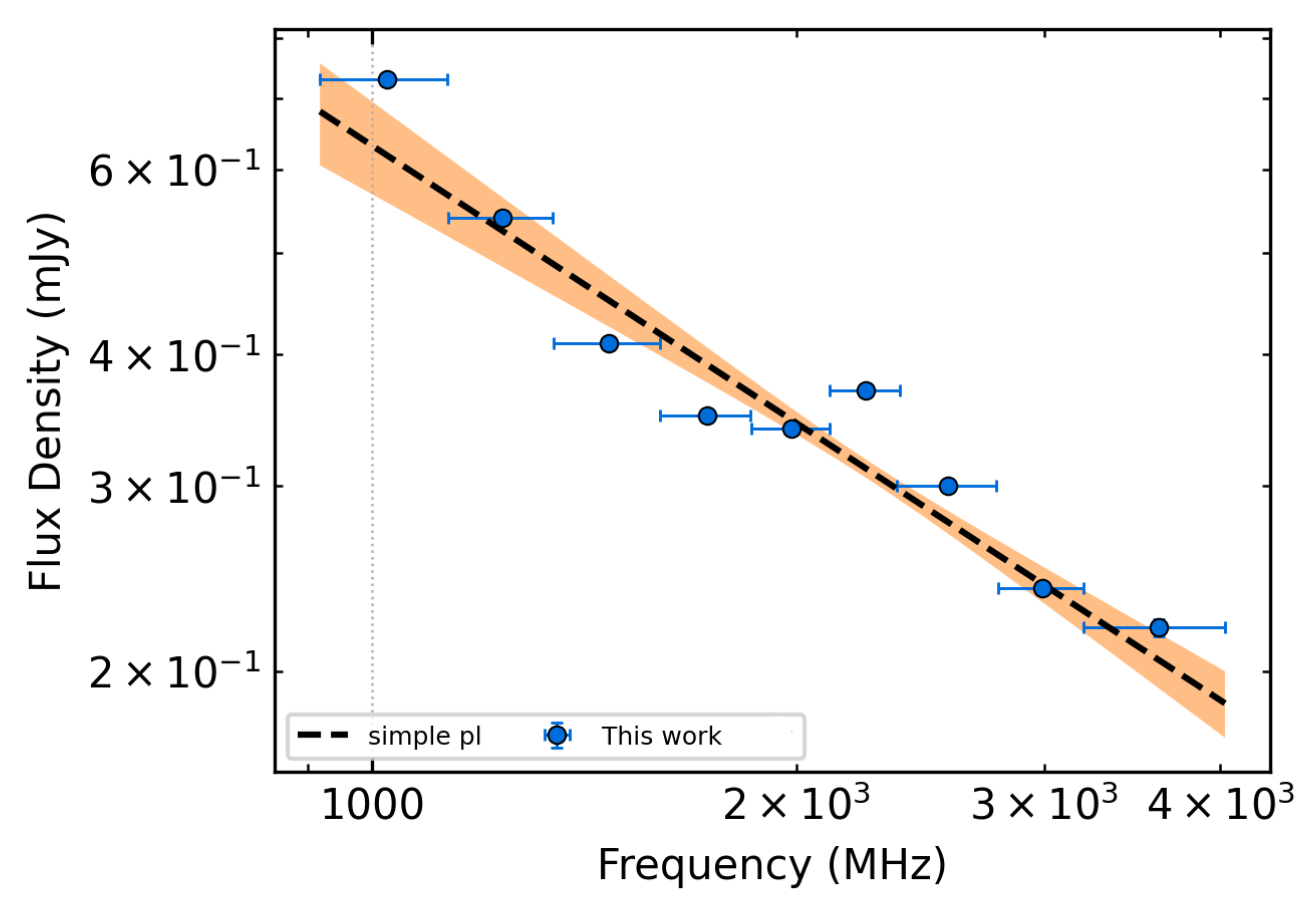} &
            \includegraphics[width=0.5\linewidth]{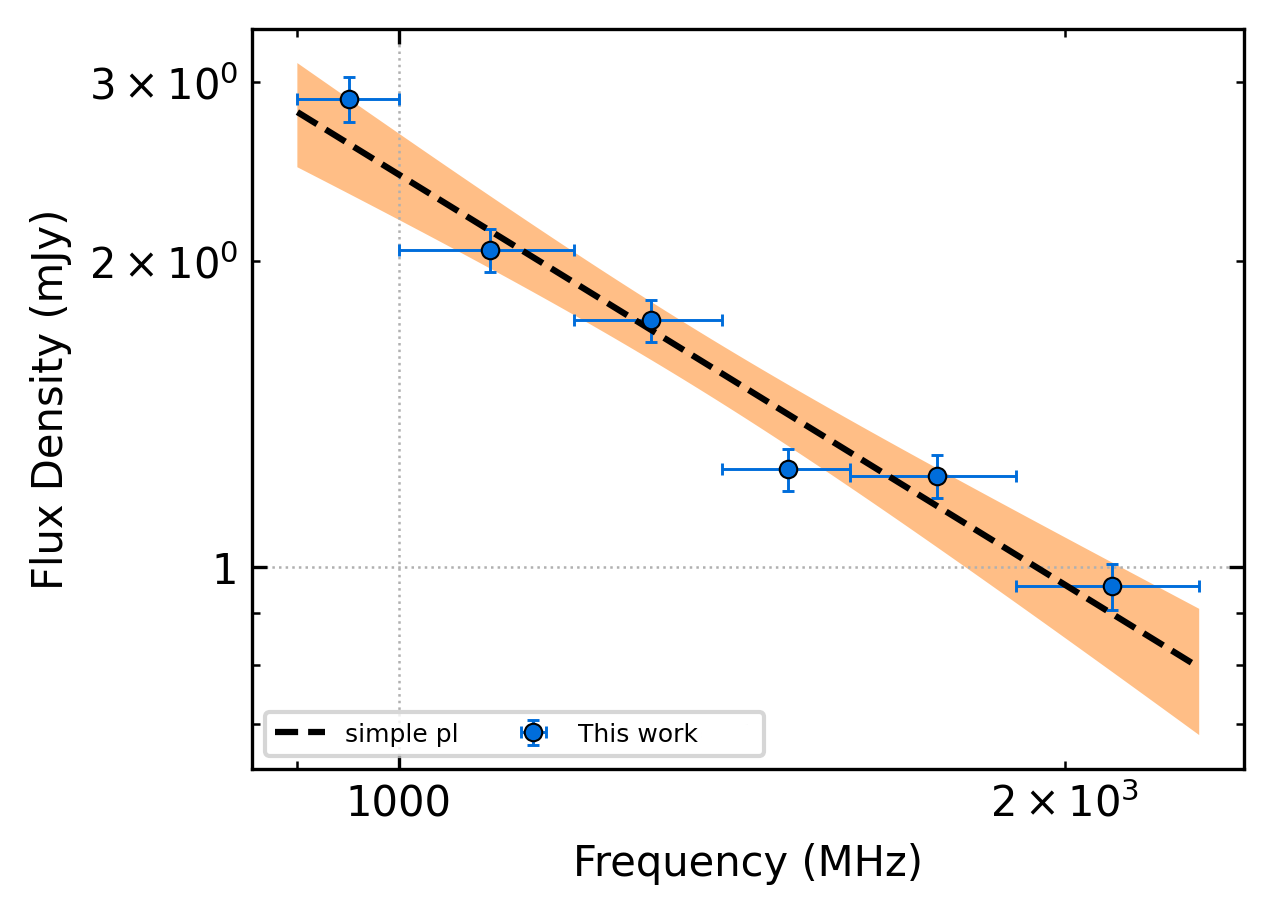}\\
            (a)Integrated pulse & (b)An example of single pulses\\
    \end{tabular}
    \caption{Spectral of PSR J1919+1745.}
    \label{1919spe}
 \end{figure}
 
It is evident that all profiles exhibit narrow double-peak characteristics. The width of the observed pulses is 0.6 ms, and it shows negligible variation with a changing frequency. The peak flux densities of the two components of the total integrated profile are similar. At lower frequencies (918-1600 MHz), the latter component dominates. At middle frequencies (1601-2368 MHz), the flux densities of both peaks are similar. Towards the higher frequencies (2369-4032 MHz), the earlier component shows a slightly larger peak flux density. Furthermore, the flux densities of both peaks exhibit a decreasing trend as the frequency increases. Using the method described in Section 3.3, we found that the spectra of both the integrated and individual pulses can be characterized by a simple power law, which takes the following form:
\begin{equation}
\centering{
    S_{\nu} = c(\frac{\nu}{\nu_0})^{\alpha},
    	\label{eq:simple}\
}
\end{equation}
where $\alpha$ is the spectral index, $\nu_0$ is the center frequency and c is a constant. 

Figure~\ref{1919spe} (a) shows that the spectral index of the integrated pulses is -0.9, which is flatter compared to the average spectral index of most pulsars \citep[-1.60$\pm$0.03,][]{2018MNRAS.473.4436J}. A spectral fitting analysis was conducted on a sample consisting of 49 individual pulses with high signal-to-noise ratios, and all the results followed the simple power law. The average spectral index of 49 single pulses is -1.0, ranging from -3.5 to 0.01. An example of single-pulse spectrum is shown in Figure~\ref{1919spe} (b).

\subsubsection{PSR J1909+0641}

During the observation period of 2 hours, we identified 116 burst pulses, corresponding to a burst rate of 58 $h^{-1}$. 
\begin{figure}[h]
\centering
\includegraphics[width=9cm]{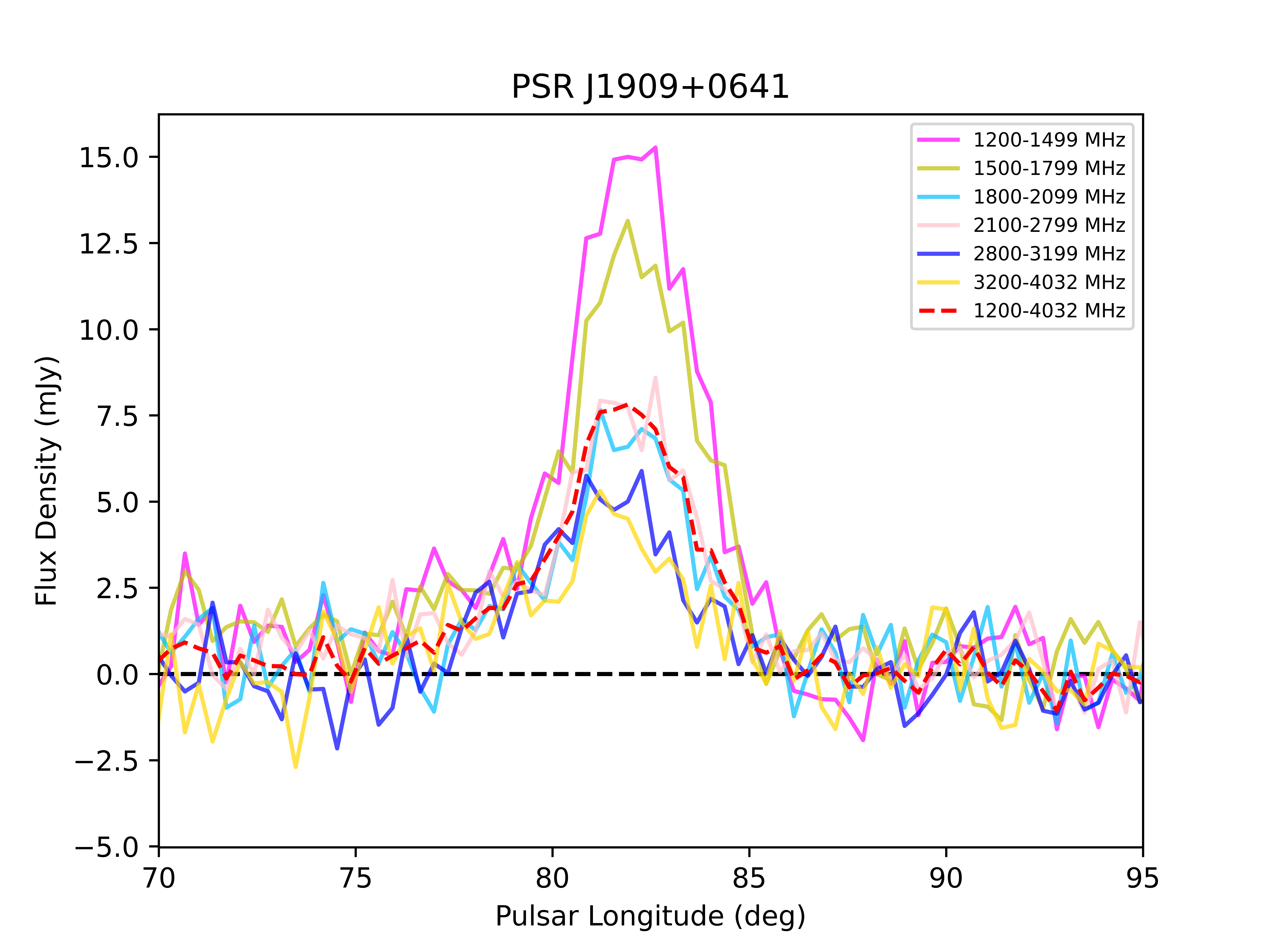}
\caption{Integrated profiles of J1909+0641 across 6 frequency sub-bands.
\label{1909-6} }
\end{figure}
\begin{figure}[thbp!]
    \centering
    \begin{tabular}{@{\extracolsep{\fill}}c@{}c@{\extracolsep{\fill}}}
            \includegraphics[width=0.5\linewidth]{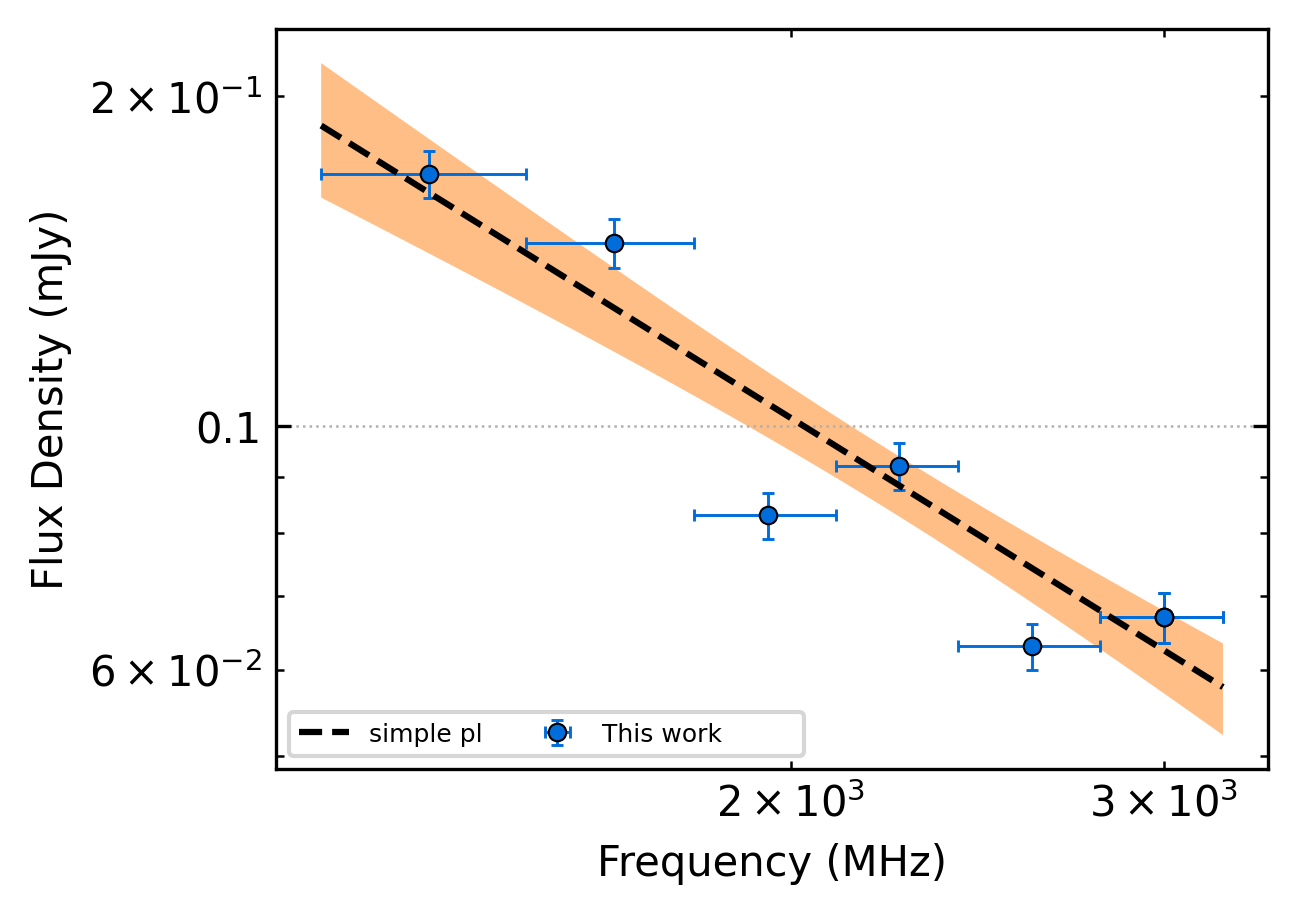} &
            \includegraphics[width=0.5\linewidth]{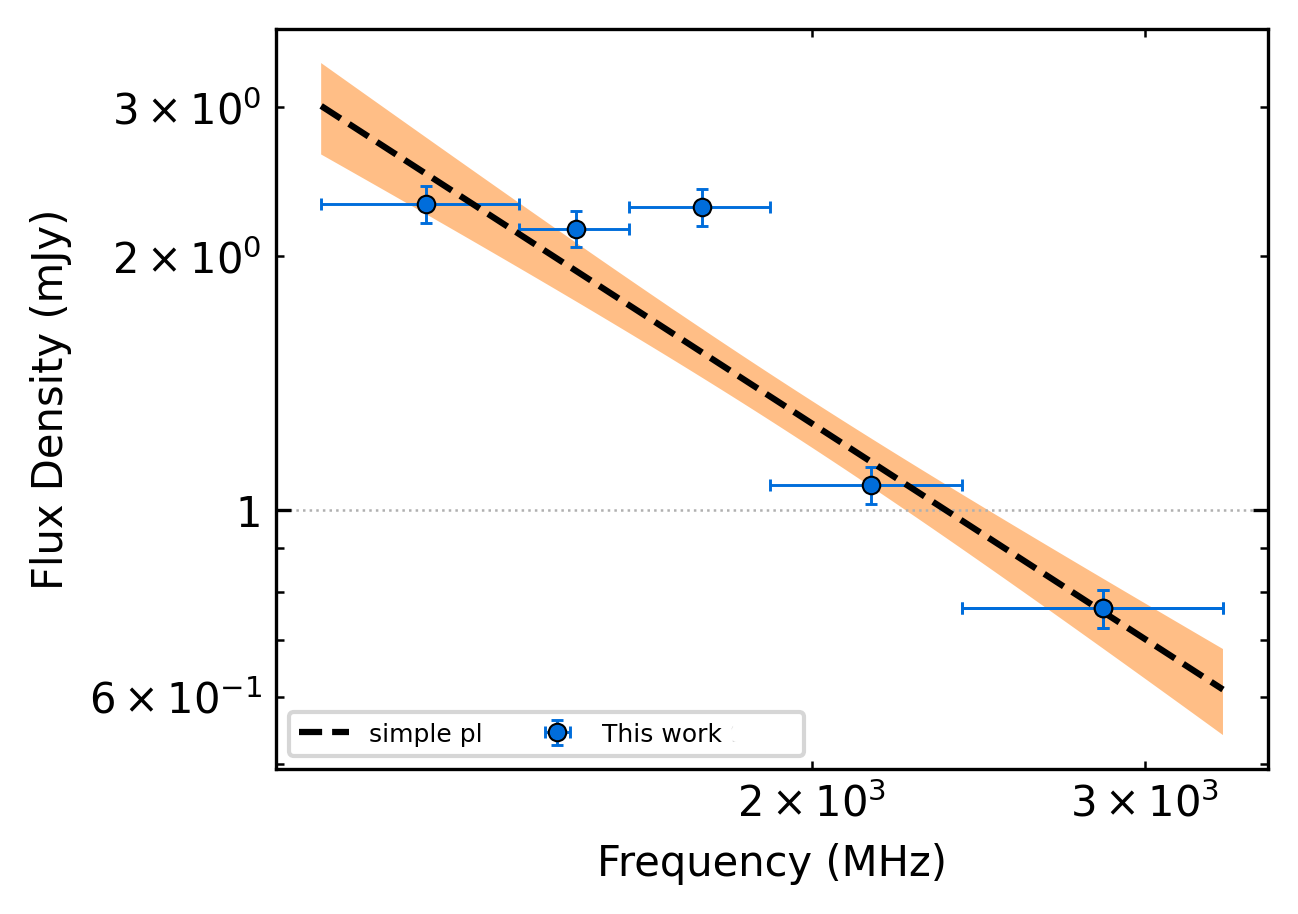}\\
            (a)Integrated pulse & (b)An example of single pulses\\
    \end{tabular}
    \caption{Spectral of PSR J1909+0641.}
    \label{1909spe}
 \end{figure} The integrated profiles, encompassing the entire frequency band and six sub-bands, exhibit a distinct narrow feature, as depicted in Figure~\ref{1909-6}. The pulse width across various frequency bands is $\sim$0.08 ms, and the peak flux density aligns with the general trend of decreasing with increasing frequency.

We performed spectral analysis on the integrated pulse as well as on seven distinct single pulses.  
It is noteworthy that all spectra exhibited a simple power-law distribution. As shown in Figure~\ref{1909spe} (a), the spectral index of the integrated pulse is -1.2, which is relatively flatter compared to the average spectral index of pulsars. Furthermore, the average spectral index of individual pulses is nearly -1.0, spanning from -1.5 to 0.05, as presented in Figure~\ref{1909spe} (b).

\subsubsection{PSR J0628+0909}
We detected 61 individual pulses in 1 hour, corresponding to a burst rate of 60 $h^{-1}$. The pulse width is 0.11 ms. The data are split into 7 sub-bands. We injected the flux densities of the 7 sub-bands into the {\sc pulsar\_spectra} flux density catalogue. The spectrum of integrated pulse can be characterized as a simple power law with a spectral index of -1.3, which is relatively flatter than the average spectral index of normal pulsars.
\begin{figure}[h]
    \centering

    \begin{minipage}[t]{1.0\linewidth}
    \centering
        \begin{tabular}{@{\extracolsep{\fill}}c@{}c@{}c@{\extracolsep{\fill}}}
            \includegraphics[width=0.315\linewidth]{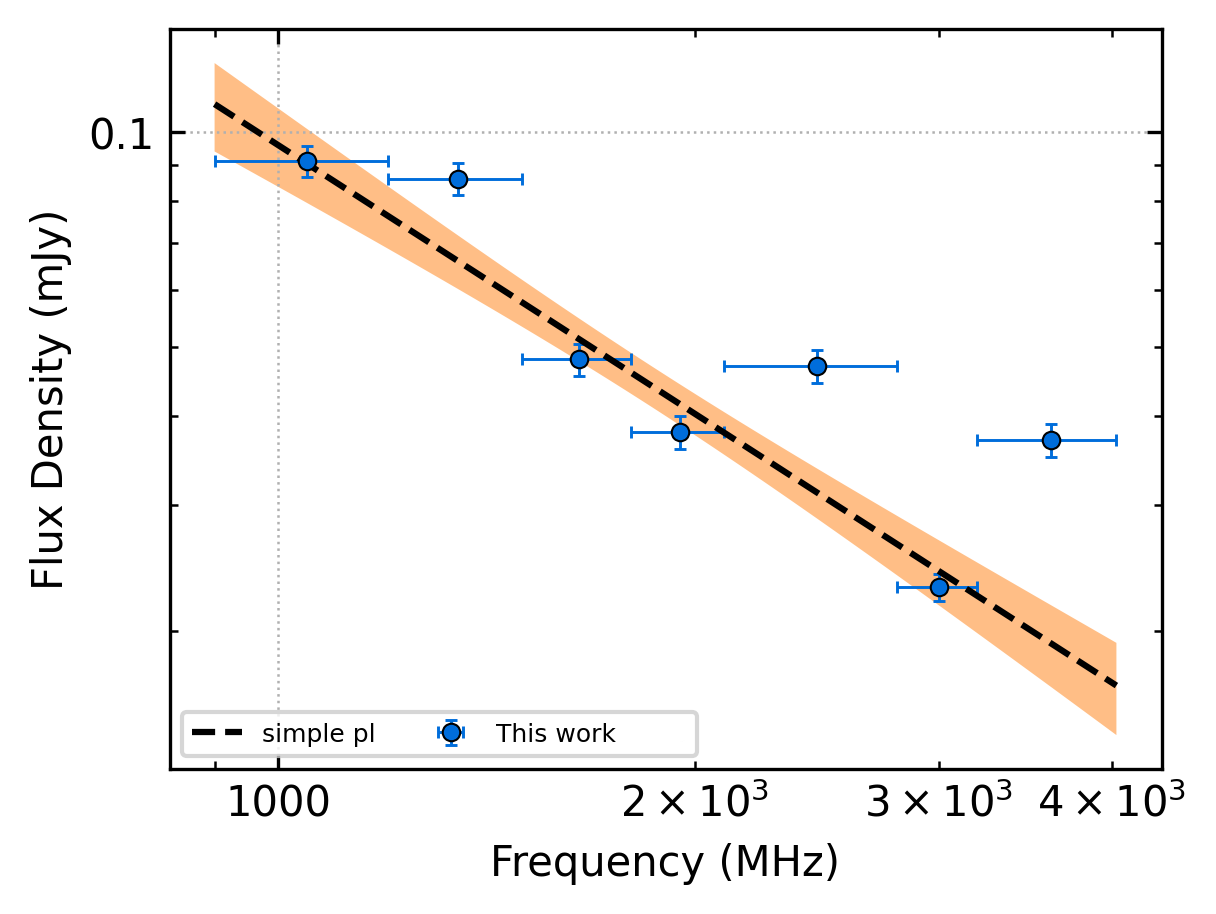}&
            \includegraphics[width=0.335\linewidth]{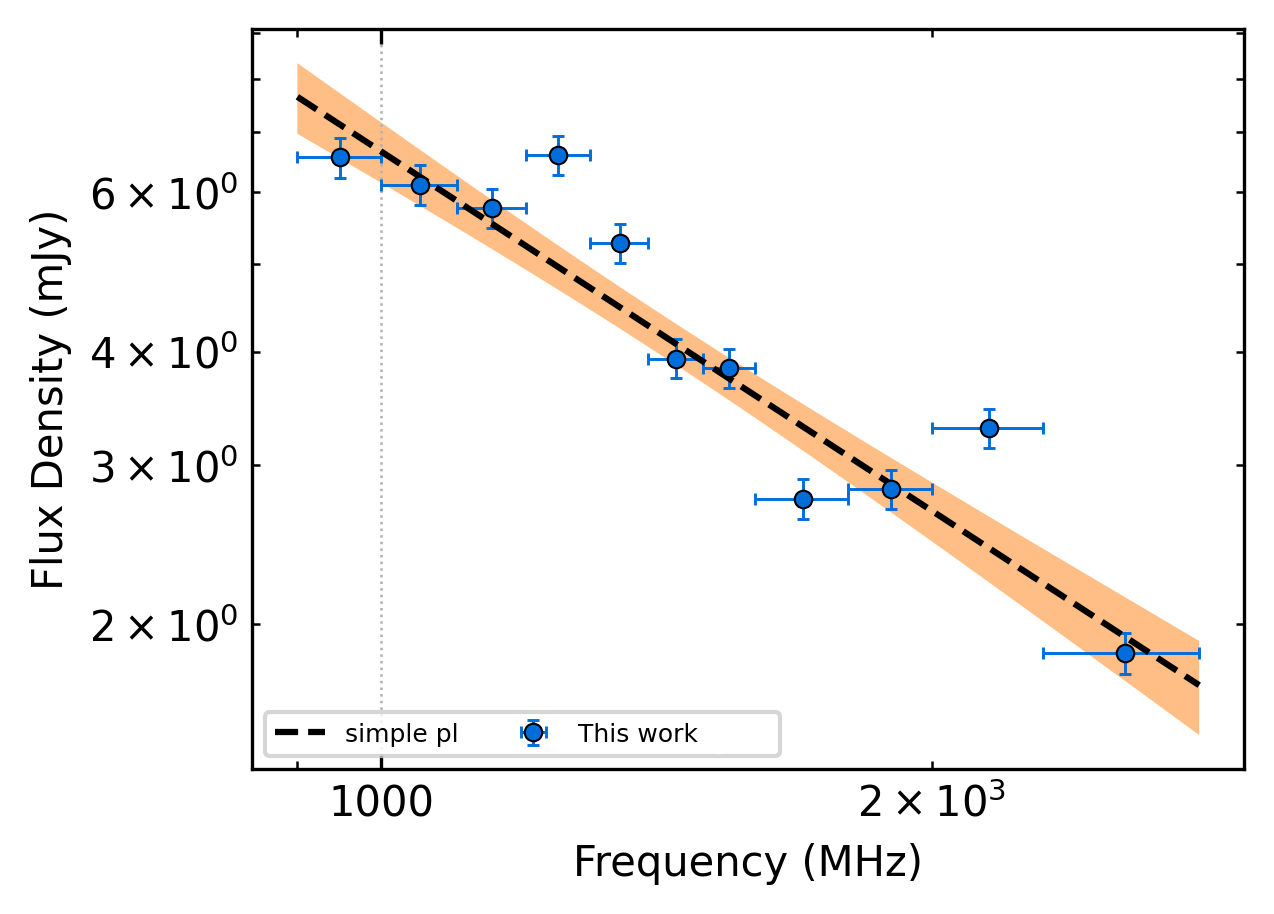}&
            \includegraphics[width=0.335\linewidth]{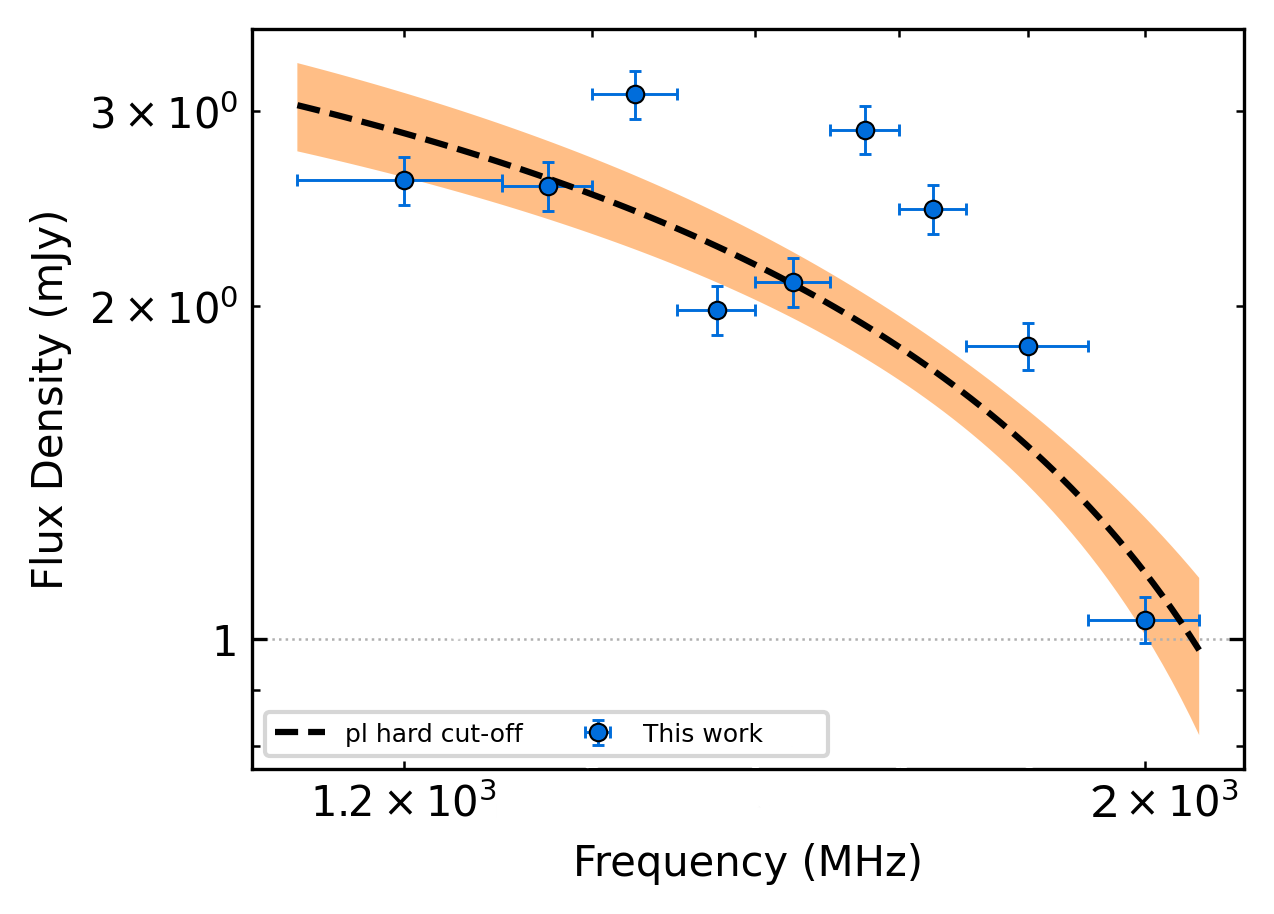}\\
            (a)Integrated pulse&(b)An example of single pulses&(c)An exceptional single Pulse\\
        \end{tabular}
    \end{minipage}
    
    \caption{Spectral of PSR J0628+0909.}
    \label{0628spe}
 \end{figure} Furthermore, we implemented spectral fitting on 22 high signal-to-noise ratio single pulses using the same approach. The vast majority of the single pulse spectrum also follows a simple power law, as presented in Figure~\ref{0628spe} (a)(b), with an average spectral index of -1.0 and a range of -1.9 to 0.5. Notably, there is an intriguing single pulse which can be divided into 11 sub-bands, and its spectrum can be modeled as a high-frequency cut-off power law, as shown in Figure~\ref{0628spe} (c), which takes the form:
\begin{equation}
    S_\nu = c\left(\frac{\nu}{\nu_0}\right)^\alpha\left(1-\frac{\nu}{\nu_c}\right),\nu\textless\nu_c,
\end{equation}
where $\alpha$ is the spectral index, $\nu_c$ is the cut-off frequency and c is a constant. This model exhibits a spectrum steepening or interruption at high frequencies. A possible explanation suggests that radiation originates in the inner (polar) gap, where electrons are accelerated in an electric field increasing from zero at the star's surface. In this process, electron acceleration reaches a maximum and decreases to zero as their velocity approaches the speed of light. All emitted power is concentrated within the radio frequency band \citep{2013Ap&SS.345..169K}.

\subsubsection{ The other RRATs with search mode}

For the rest of RRATs observed with the UWL receiver, we could not create reliable profiles. Therefore, we are unable to carry out subsequent processing and analysis, as we have done for the previous three RRATs.

Limits on the flux density of a single pulse can be described by the equation \citep{2021arXiv210604821T}:
\begin{equation}
\centering{
    S_{lim} = \frac{\sigma S/N_{min}T_{sys}}{G\sqrt{\Delta \nu
    N_pt_{obs}}},
    	\label{eq:5}\
}     
\end{equation}
where $T_{sys}$=22 K is the system temperature, G=1.8 K $Jy^{-1}$ is the antenna gain for the UWL receiver of Parkes telescope ,$t_{obs}$ is the observation time, $\sigma$ is the root-mean-square (rms) noise level and $\Delta\nu$ is the full 3300 MHz bandwidth. As for periodic signals, equation~\ref{eq:5} should be multiplied by $\sqrt{\frac{\delta}{1-\delta}}$, where $\delta$ is the duty cycle. Due to the absence of a measured pulse width for PSR J1850+15 in previous investigations, we assume the pulse width to be 1 ms and a flat spectrum, our non-detection of signal with S/N above 7 puts a fluence limitation of $\sim$13 mJy ms. Detailed parameters regarding the limitation of flux density and fluence for the 5 RRATs are listed in Table~\ref{tab2}.

\begin{table}[h!]
\bc
\centering
\begin{minipage}[]{100mm}
\caption[]{Summary of the flux density and fluence limits of the single pulses and periodicity search of 5 RRATs with Parkes UWL receiver.\label{tab2}}\end{minipage}
\setlength{\tabcolsep}{1pt}
\small
 \begin{tabular}{lcccccccccccc}
  \hline\noalign{\smallskip}

PSR Name& Period & $T_{obs}$ & $\sigma$ & Assuming Width &Width & Flux Density Limit ($7\sigma$) &Fluence Limit ($7\sigma$)  \\
 &(s)&(s)&(rms)&Single Pulses(ms)&Periodic Signal&Single Pulse/Periodic Signal&Single Pulse/Periodic Signal\\
\hline
J0627+16     &  2.2 &2442&30.0&0.03&0.3&21.4mJy/7.5$\mu$Jy&0.6mJy ms/2.3$\mu$Jy ms\\
J1913+1330         & 0.9&3168&34.7&0.2&2.0&38.0mJy/30.2$\mu$Jy&7.6mJy ms/60.4$\mu$Jy ms\\
J1928+15        & 0.4&1210&12.6&0.5&5.0&21.0mJy/42.9$\mu$Jy&10.5mJy ms/214.5$\mu$Jy ms\\
J1946+24          & 4.7&1166&21.8&0.4&4.0&10.6mJy/19.6$\mu$Jy&4.2mJy ms/78.3$\mu$Jy ms\\

 J1850+15&1.4&2554&32.4&0.1&1.0&29.0mJy/18.1$\mu$Jy&2.9mJy ms/18.1$\mu$Jy ms\\

  \noalign{\smallskip}\hline
\end{tabular}
\ec

\tablecomments{\textwidth}{The pulse widths for the periodic signals of PSRs J0627+16, J1913+1330, J1928+15, and J1946+24 are from \cite{2009ApJ...703.2259D} and \cite{2009MNRAS.400.1431M}. As for PSR J1850+15, its pulse width was not found in previous works. Therefore, it is assumed to be 1 ms. Additionally, it is assumed that the pulse widths of all individual pulses are one-tenth of the pulse width of periodic pulses.}
\end{table}
\subsection{Fold-mode data}
\subsubsection{PSR J1709-43}
\begin{figure}[h]
\centering
\includegraphics[width=9cm]{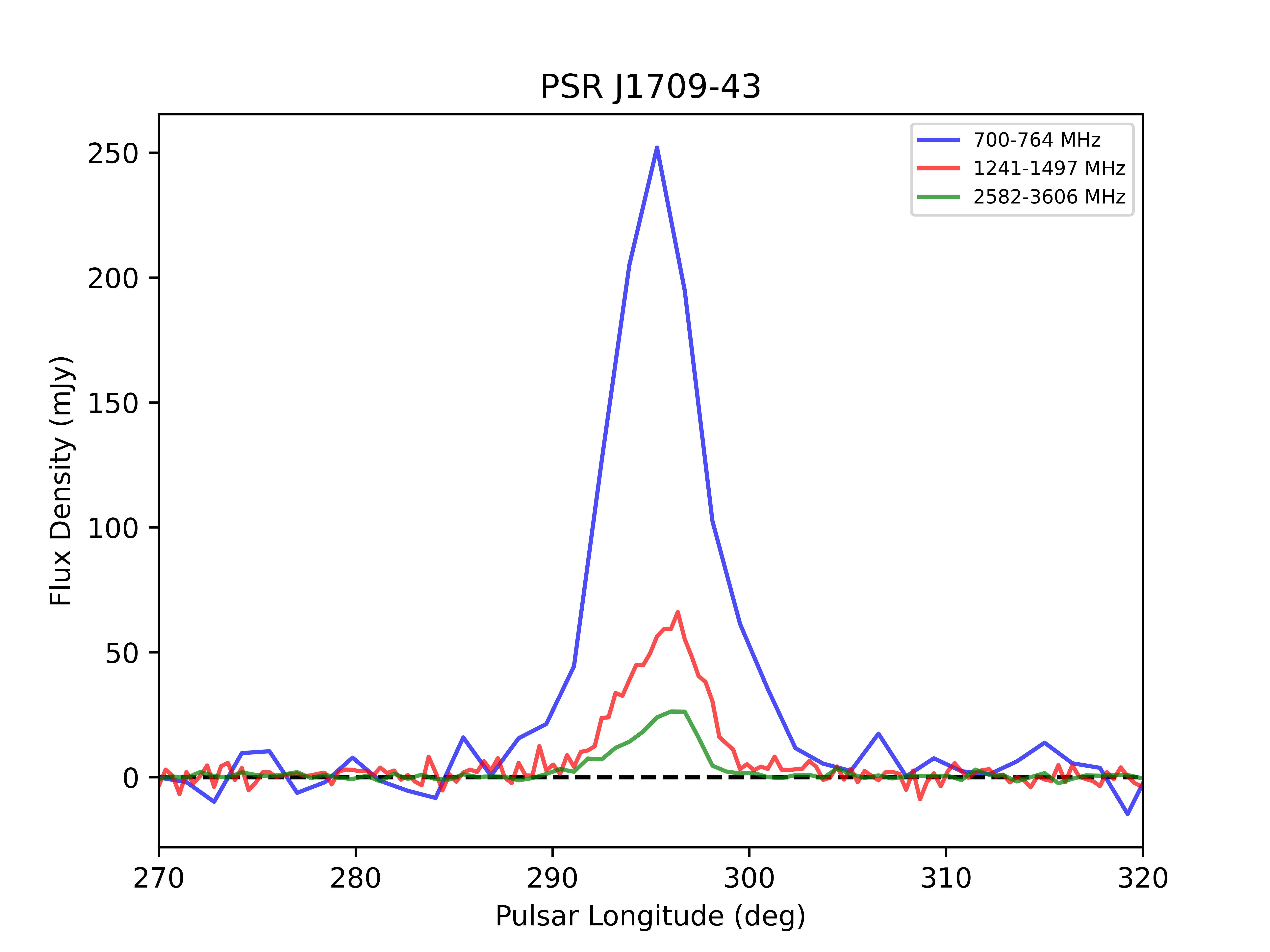}

\caption{Integrated profiles of PSR J1709-43 across 3 frequency sub-bands.\label{1709} }
\end{figure}

We collected observational data spanning 7.2 years, from MJD 55537 to 58168. A total of 103 observations were conducted, all in a folded mode. Pulse profiles were detected in 41 observations, constituting $\sim$40$\%$ of the overall data set. The majority exhibits continuous pulse signals across all sub-integrations, while only 7 observations experienced nulling. Notably, the highest detection rate was obtained at 1400 MHz, accounting for $\sim$56$\%$ of the observations. Three total intensity pulse profiles of PSR J1709-43 at 732 MHz, 1369 MHz, and 3094 MHz are in good agreement which exhibit a narrow single-peaked structure, as presented in Figure ~\ref{1709}. 
These profiles follow the trends of decreasing width and peak flux density with increasing frequency. In addition to the analysis presented in Section 3, we performed a timing analysis for PSR J1709-43, which enabled us to determine its rotating period and the first derivative of the rotating period. The full timing solutions are listed in Table~\ref{tab3}.

\begin{table}[h]
\centering
\caption{Ephemeris for PSR J1709-43.}\label{tab3}
\begin{tabular}{lc}
\hline

Parameter &Value\\
\hline

Pulsar Name&PSR J1709-43\\
Right Ascension (RA, J2000)        & 17:09:46 \\
Declination (Dec, J2000)         & -43:54:32 \\
Reference epoch (PEPOCH, MJD)           & 56800\\
Data Span (MJD)         & 55537-58168  \\ 
Dispersion measure (DM)       & 228 pc $cm^{-3}$\\
Rotation frequency ($F_0$) & 1.11499999997 Hz\\
Frequency derivative ($F_1$)& $-3.0038\times10^{-14}$ $s^{-2}$\\
Rotation period ($P_0$)&0.8968609868 s \\
Period derivative ($P_1$) & $2.4162\times10^{-14}$ $ss^{-1}$ \\
Surface magnetic field ($B_s$) &4.71$\times$$10^{12}$ G\\
Characteristic age ($\tau$) & 0.59 Myr\\

\hline

\end{tabular}
\end{table}

\subsubsection{PSR J1649-4653}
Similar to PSR J1709-43, we have collected data for PSR J1649-4653 covering 11 years, from MJD 54220 to 58222, with a total observation time of 10 hours. All the observations were carried out in the folded mode. Out of these, 65 observations ($\sim$51$\%$) detected the burst pulse profiles. Figure 7 illustrates the variation of peak flux density of J1649-4653 with time. The flux density of PSR J1649-4653 is found to be significantly weak, with a value of 0.14$\pm$0.02 mJy for the detectable pulse profiles.

\begin{figure}[h]

\centering
\includegraphics[width=10cm]{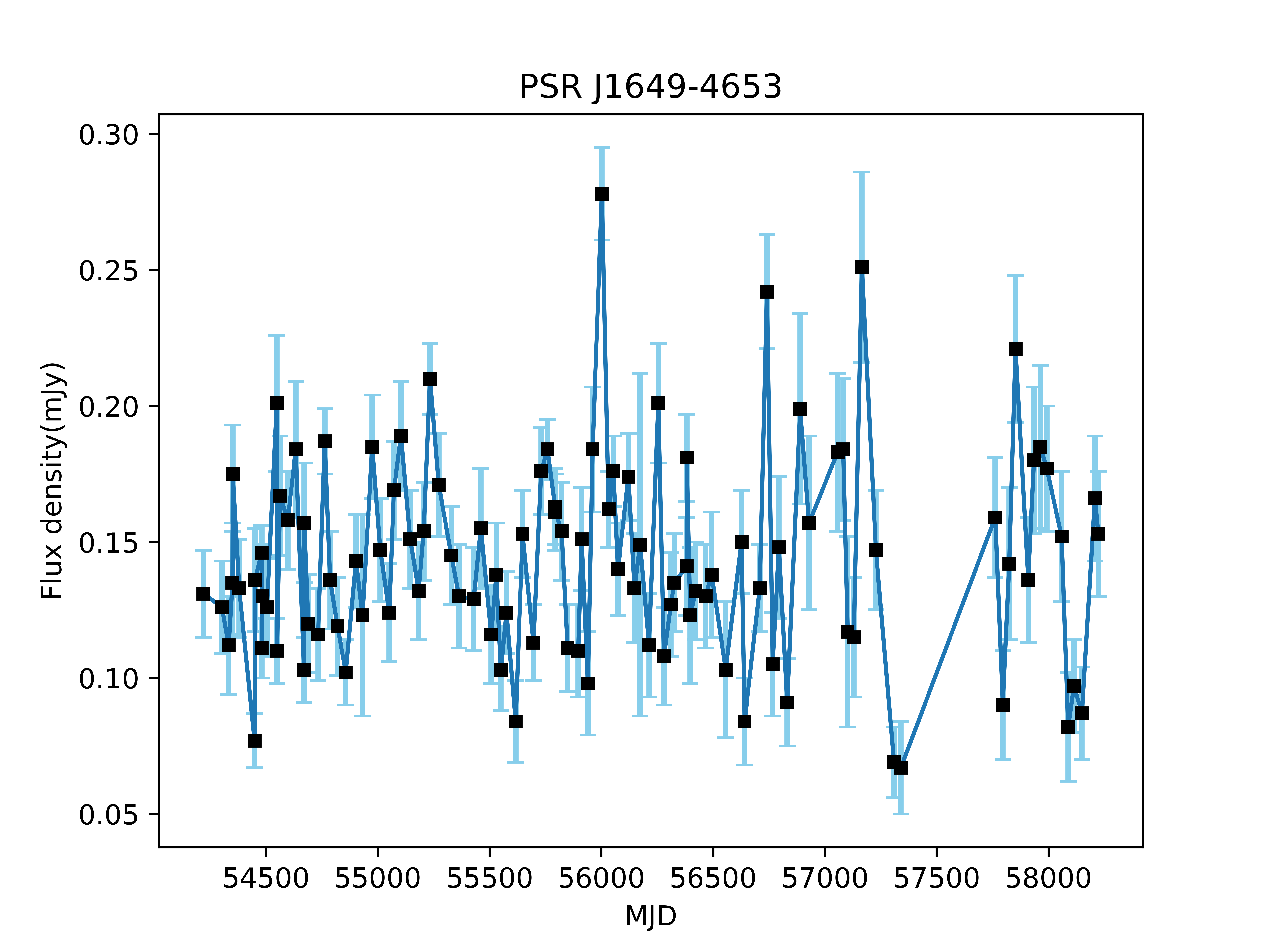}

\caption{Flux density as a function of MJD for PSR J1649-4653.\label{1649} }
\end{figure}

\section{Discussion}

We calculated the burst rates for three RRATs, PSRs J1919+1745, J1909+0641, and J0628+0909. These rates were then compared with results from the Arecibo telescope \citep{2009ApJ...703.2259D} and the Five-hundred-meter Aperture Spherical radio Telescope \citep[FAST,][]{2023MNRAS.518.1418H}, as detailed in Table~\ref{tab4}. Notably, J1919+1745 exhibited outstanding behavior with an observed burst rate of 847 $h^{-1}$, which significantly exceeds the previously reported rate of 320 $h^{-1}$ from the Arecibo Telescope. Furthermore, the burst pulses constituted $\sim$50$\%$ of all the observed individual pulses. Consequently, we suggest that PSR J1919+1745 may be reclassified from a RRAT to a nulling pulsar. 
For PSR J1909+0641, the observed burst rate was 58 $h^{-1}$, relatively lower than the rate of 67 $h^{-1}$ observed at Arecibo telescope. For PSR J0628+0909, the burst rate observed in this study was 60 $h^{-1}$. In comparison, the Arecibo reported a rate of 141 $h^{-1}$ (S/N$\textgreater$5), and the FAST reported a rate of 270 $h^{-1}$ (S/N$\textgreater$7). We collected data from 103 observations for PSR J1709-43 in fold mode. Among these, pulse profiles is detected in 41 observations ($\sim$40\%). Pulse signals can be seen across all sub-integrations for most of them, with only 7 observations experiencing nulling. 128 observations were collected  for PSR J1649-4653 in fold mode, pulse emissions were detected for approximately 51\% (65 observations).  
Our analysis suggests that the variations in burst rates can be attributed to three primary factors. Firstly, variations in telescope sensitivity and observation bandwidth play a vital role. 
If telescopes with higher signal-to-noise ratios are employed for observations, it is possible that a greater number of burst pulses could be detected. The different burst rates observed in PSR J0628+0909 may be attributed to the sensitivity of different telescopes. The FAST telescope demonstrates the highest sensitivity, corresponding to the highest burst rate. Secondly, variations in the duration of observations can exert an influence. The observation durations of the FAST and the Arecibo telescopes were comparatively shorter than that of the Parkes telescope. Lastly, it is worth noting that the burst rates of RRATs may exhibit temporal evolution.

\begin{table}[h]
\bc
\centering 
\caption{Comparison of burst rates detected of three telescopes.}\label{tab4}
\begin{tabular}{lccccccccc}
\hline

PSR&\multicolumn{3}{c}{Burst Rate}&\multicolumn{3}{c}{Frequency}&\multicolumn{3}{c}{Duration}\\

\multicolumn{1}{c}{}&Parkes&Arecibo&FAST&Parkes&Arecibo&FAST&Parkes&Arecibo&FAST \\
\multicolumn{1}{c}{}&($h^{-1}$)&($h^{-1}$)&($h^{-1}$)&(MHz)&(MHz)&(MHz)&(h)&(h)&(h)
\cr
\hline
J1919+1745&847&320 &$-$&732-4032 &1390-1490 &$-$&1.0 &0.11&$-$ \\
J1909+0641&58 &67 &$-$&732-4032  &1390-1490 &$-$&2.0 &0.15 &$-$\\
J0628+0909&60&141 &270&732-4032&1390-1490 &1000-1500&1.0&0.30&0.48\\

\hline
\end{tabular}
\ec
\tablecomments{\textwidth}{The observational data in this study were acquired using the Parkes telescope. The Arecibo data were sourced from \cite{2009ApJ...703.2259D}, while the FAST data were obtained from \cite{2023MNRAS.518.1418H}.}

\end{table}

We analyzed the integrated profiles spanning multiple frequency sub-bands for PSRs J1919+1745, J1909+0641, and J1709-43. A notable trend was discerned: the peak flux densities exhibited a diminishing trend with increasing frequencies, consistent with the previous results. 

Subsequently, we carried out an analysis and fitting of the flux density in various frequency sub-bands for both integrated and single pulses. This analysis was facilitated using an automated spectral fitting software, {\sc pulsar\_spectra}.
We present the spectral analyses of the integrated pulses emanating from three RRATs, PSRs J1919+1745, J1909+0641, and J0628+0909. All of them consistently follow a simple power law. The integrated results are consistent with previous work, which presented that 79\% of pulsars have spectra with a simple power law \citep{2018MNRAS.473.4436J}. 

However, there is a dearth of literature available for comparing single-pulse spectral index with our results. It's remarkable that only one single-pulse spectral fitting of PSR J0628+0909 exhibits a high-frequency cut-off power-law spectrum with a cut-off at 1625 MHz. 
Complex spectral fitting typically requires a substantial number of flux density at different frequencies.  
However, our current limitations lie in acquiring a significant volume of high-quality single-pulse spectral flux density data. This may lead to a predominant trend in our fitting results, favoring a simple power-law model. \cite{2018MNRAS.473.4436J} summerized that there are 3 scenarios for deviations from a simple power-law model in pulsars can be considered: an environmental origin of the observed spectral features, absorption processes in the magnetosphere, and non-generic emission characteristics. The calculated mean of the single-pulse average spectral indices for the aforementioned three RRATs, as presented in Table~\ref{tab5}.
\begin{table}[h]

\centering
\caption{Spectral indices of 3 RRATs}\label{tab5}
\begin{tabular}{lcccc}
\hline

PSR&Integrated Pulse&\multicolumn{3}{c}{Single Pulse}\\
\multicolumn{1}{c}{}&&Mean&Standard Deviation&Range\\
\hline
J1919+1745&-0.9&-1.0&1.0&-3.5 - 0.2\\
J1909+0641&-1.2&-1.0&0.5&-1.5 - 0.05\\
J0628+0909&-1.0&-1.3&0.7&-2.0 - 0.5\\

\hline
\end{tabular}

\end{table}
Nevertheless, there is a dearth of literature available for comparing single-pulse spectral indices with our results. Previous studies by 
\cite{2018ApJ...866..152S} reported that the mean spectral indices 
for PSRs J1819-1458, J1317-5759, and J1913+1330 are -1.1, -0.6, and -1.2 respectively. The range of single-pulse spectral indices they obtained is from -7 to +4.
\cite{2022ApJ...940L..21X} reported that the mean spectral index for single pulses is -3.2 for PSR J0139+3336 and showed an extensive range from -11.85 to +3.83. 
The frequency ranges for these two observations were relatively narrow, spanning 288 MHz and 500 MHz, respectively.
\cite{2019PASA...36...34M} reported that the single-pulse mean spectral index for PSR J2335-0530 is -2.2, ranging from -2.8 to -1.5. 
The observational data were obtained from two frequency bands: one centered at 154.24MHz with a bandwidth of 30.72MHz, and the other centered at 1396MHz with a bandwidth of 256MHz. Consequently, the observation covered a broad frequency range.
Our results exhibit relatively lower scatter compared to the results from \cite{2018ApJ...866..152S} and \cite{2022ApJ...940L..21X}, which may be attributed to the broader frequency range we measured (3300 MHz) and the longer observation duration.

\section{Summary}
We performed an analysis of the emission properties for 10 RRATs, obtaining burst rates for 3 RRATs and integrated profiles across multiple frequency bands for 3 RRATs. We also conducted an in-depth wideband analysis of the integrated pulse and single-pulse spectral characteristics using the {\sc pulsar\_spectra} software package. Our findings indicate that the vast majority of these characteristics align with a simple power-law model, featuring relatively flat spectral indices and a comparatively small scatter in single-pulse spectral indices. For five additional RRATs observed with the UWL receiver, we provide their fluence and flux density upper limits. Furthermore, we derived the timing solution for PSR J1709-43. 
Based on the classification of 76 recently discovered RRATs by the FAST telescope in the previous work of \cite{2023arXiv230317279Z}, we suggest that PSRs J1919+1745, J1709-43 and J1649-465 may be nulling pulsars or weak pulsars with sparse strong pulses.

\begin{acknowledgements}

The Parkes radio telescope is part of the Australia Telescope National Facility, which is funded by the Australian Government funds operation as a National Facility managed by CSIRO. This paper includes archived data obtained through the CSIRO Data Access Portal \footnote{http://data.csiro.au}. This work is supported by the Major Science and Technology Program of Xinjiang Uygur Autonomous Region (grant no. 2022A03013-4), the Zhejiang Provincial Natural Science Foundation of China (grant no. LY23A030001), the National SKA Program of China (grant no. 2020SKA0120100, 2022YFC2205201, 2020SKA0120200), the Natural Science Foundation of Xinjiang Uygur Autonomous Region (grant no. 2022D01D85), the National Natural Science Foundation of China (grant no. 12041304,12273100, 12041303), and the West Light Foundation of Chinese Academy of Sciences (grant no. WLFC 2021-XBQNXZ-027), and the open program of the Key Laboratory of Xinjiang Uygur Autonomous Region (grant no. 2020D04049).

\end{acknowledgements}

	\nocite{*}
	
	\bibliography{Ren.bib} 
	
	\bibliographystyle{raa} 

\end{document}